\documentclass[a4paper]{article}
\usepackage[margin=1in]{geometry}
\usepackage{authblk}
\usepackage{amsmath}

\usepackage{graphicx}

\usepackage{anyfontsize}

\usepackage{titlesec}
\titleformat{\chapter}[display]{\huge\bfseries}{\chaptertitlename~\thechapter}{20pt}{\Huge}
\titlespacing{\chapter}{0pt}{50pt}{40pt}
\title{{Energy Deposition along The Helical Undulator at ILC-250GeV*}}

\author[1,2,4]{Khaled Alharbi} 
\author[2]{Sabine Riemann} 
\author[1,3]{Gudrid Moortgat-Pick} 
\author[4]{Ayash Alrashdi}

\affil[1]{\textit{University of Hamburg, Luruper Chaussee 149, D-22761 Hamburg, Germany}}
\affil[2]{\textit{Deutsches Elektronen-Synchrotron (DESY), Platanenallee 6, D-15738 Zeuthen, Germany}}
\affil[3]{\textit{Deutsches Elektronen-Synchrotron (DESY), Notkestrasse 85, D-22607 Hamburg, Germany}}
\affil[4]{\textit{King Abdul-Aziz City for Science and Technology (KACST), Riyadh, Kingdom of Saudi Arabia}}

\date{}
\begin{document}\maketitle

\begin{abstract}

\fontsize{13}{13}\selectfont
The positron source of the International Linear Collider is based on a superconducting helical undulator passed by the high-energy electron beam to generate photons which hit a conversion target. Since the photons are circularly polarized the resulting positron beam is polarized. At ILC250, the full undulator is needed to produce the required number of positrons. To keep the power deposition in the undulator walls below the acceptable limit of 1W/m, photon masks must be inserted in the undulator line. The photon mask design requires a detailed study of the power deposition in the walls and masks. This paper describes the power deposition in the undulator wall due to synchrotron radiation.

\end{abstract}

\fontsize{12}{22}\selectfont
\section{Introduction}

The baseline positron source of the International Linear Collider (ILC) uses a helical undulator. The electron beam passes a long super-conducting (SC) helical undulator that is placed at the end of main Linear Accelerator. A high number of circularly-polarized photons is produced. These multi-MeV photons hit a thin Ti6Al4V target in order to produce longitudinally polarized positrons through the pair production mechanism \cite{adolphsen2013international}.

One great benefit of the undulator-based positron source is the generation of a longitudinally polarized positron beam  \cite {flottmann1993investigations}. Without collimation of the photon beam the degree of positron beam polarization is  $\approx 30\%$ which already improves substantially the precision of measurements.

\begin{flushleft}
\rule{200pt}{0.5pt}\\

\fontsize{10}{10}\selectfont
{*Talk presented at the International Workshop on Future Linear Colliders (LCWS2018), Arlington, Texas, 22-26 October 2018. C18-10-22.\\
khaled.alharbi@desy.de}
\end{flushleft}

However, to achieve the required luminosity with the baseline undulator, its full active length must be used. Since the opening angle of the photon beam depends on 1/$\gamma$, more photons could hit the undulator wall  \cite {malyshev2007vacuum}. However, the power deposited in the undulator walls should be kept below 1 W/m \cite {scott2008investigation}.

In this paper the energy deposition in the undulator wall is studied. In order to protect the wall, photon masks can be installed along the undulator at the position of the quadrupoles as suggested in \cite {bungau2008design}. For the mask design the distribution of the photons and their energy is crucial.

The ILC undulator parameters are shown in table \ref{table1}. The general layout of the undulator is described in [TDR]. The undulator modules have a length of 1.75m.  Always two undulator modules are mounted in one cryomodule of 4.1m lentgh. Quadrupoles are positioned every 3 cryomodules to steer and focus the electron beam through the undulator. The active length of the undulator is 231 m; the total length amounts to 320 m. The undulator aperture is 5.85mm. The undulator period is 11.5mm and the maximum B field on undulator axis is 0.86T corresponding to a maximum K value of 0.92. Here, K=0.85 was used.

\begin{table}[h]
\caption[ILC undulator parameters]{ILC undulator parameters.}
\centering 
\begin{tabular}{l*{6}{l}r} 
\hline\hline 
Parameters & Values \\ [0.5ex] 
\hline 
Centre-of-mass energy  & 250 GeV \\ 
Undulator period & 11.5 mm\\
Undulator K & 0.85 \\
Electron Number per bunch & $2\times10^{10}$ \\
Number of bunches per pulse & 1312 \\
Pulse rate &  5.0 Hz \\
Cryomodule Length & 4.1 m \\
Affective Magnet Length & 3.5 m \\
Undulator Aperture & 5.85 mm \\
Number of quadrupoles  & 23 \\ 
Quadrupole Spacing & 14.538 m\\
Quadrupole length & 1 m \\
Total active undulator length & 231 m \\
Total lattice length & 319.828 m \\

\hline 
\end{tabular}
\label{table1} 
\end{table}

\section{Synchrotron Radiation Power Incident on the Undulator Wall}

To calculate the incident power from the Synchrotron Radiation (SR) on the undulator wall, the angular distribution of the power radiated from the helical undulator must be studied.

The angular power distribution produced from a helical undulator with N periods can be calculated by using Equation \ref{ed24} \cite {kincaid1977short}:

\begin{equation}\label{ed24}
\frac{{{d}}I}{d\Omega }{(\theta)}=\frac{{2}{N}{{q}^{2}}{{\omega_{0}}}{{\gamma}^{4}}{{K}^{2}}}{{{\pi}}{{\varepsilon}_{0}}c{{(1+{{K}^{2}+}{{\gamma}^{2}}{{\theta}^{2}})}^{3}}}\sum\limits_{n=1}^{h}{{n}^{2}}{\left[{J'_n}^{2}\left(x
\right)+{{\left(\frac{\gamma\theta}{K}-\frac{n}{x}\right)}^{2}}J_{n}^{2}\left(x\right)\right]}
\end{equation}

Where:

\begin{align}
  K&=\frac{{{\lambda }_{u}}qB}{2\pi m{{c}^{2}}}
\end{align}

\begin{align}
  x&=\frac{2{n}K\theta\gamma }{1+{K}^{2}+\gamma^2\theta^2 }, 
\end{align}
\begin{align}
  {{\omega }_{0}}&=\frac{2\pi {{\beta }^{*}}c}{{{\lambda }_{u}}},
\end{align}

Where $\theta$ is the photon opening angle, $\gamma$ is the relativistic factor, and h is the number of harmonics to be included. ${\omega}_0$ is the circular frequency of the electron’s helical orbit. ${\lambda}_{u}$  is the period. 

The photons generated in the first undulator module pass the full undulator length except a part which hits the wall of modules downstream. Figure 1 shows the distribution of these photons along the undulator wall separated for the first six harmonics and in total. In the beginning of the undulator the contribution of photons from the 1st harmonic is order of magnitudes larger than that from higher harmonics. The incident photon number from higher harmonics is increasing along the undulator length. At the end of the undulator, the incident photon number from the second and third harmonics are close to that from the first harmonic. Figure 2 shows the average photon energy for each of these harmonics.

\begin{figure}[h]
\centering
\includegraphics[scale=0.6]{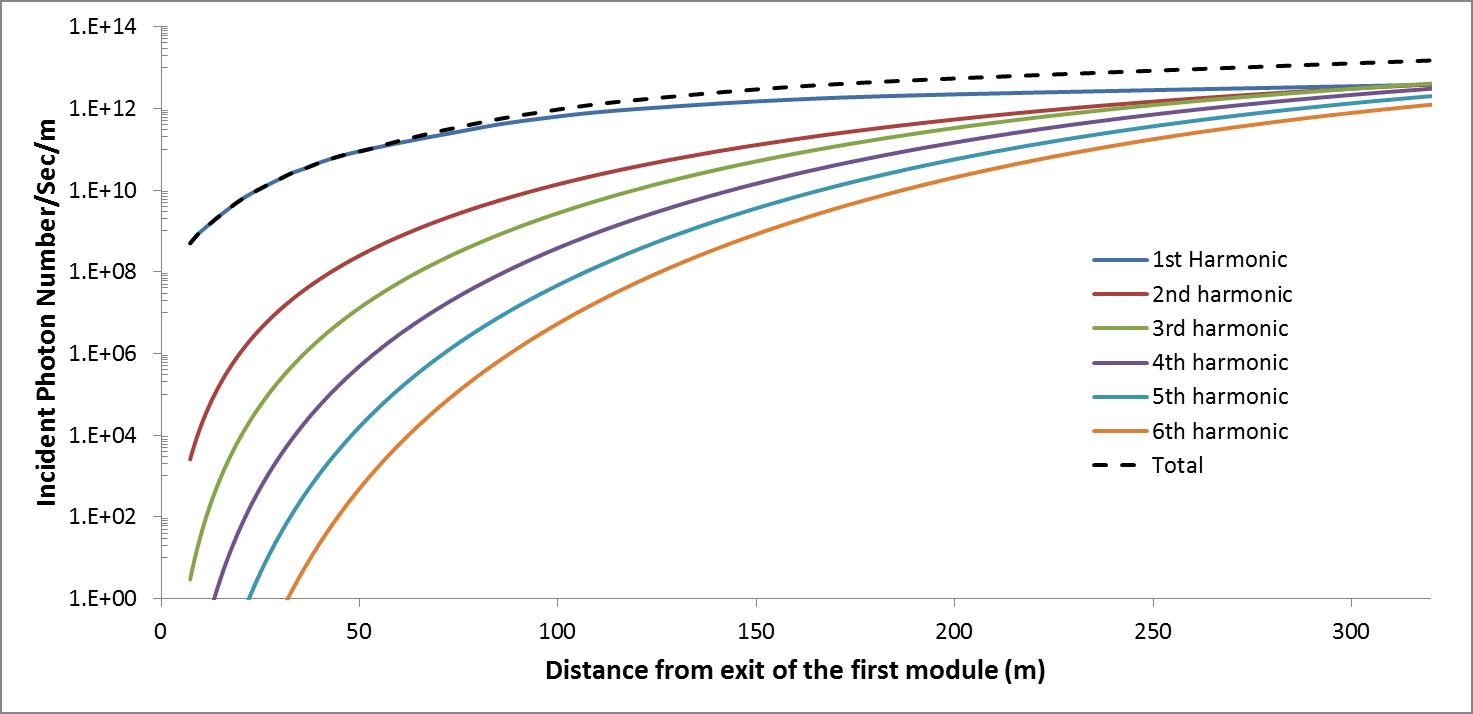}
\caption{Number of photons from the first module incident on the undulator walls.}
\label{fig:graph1}
\end{figure}

\begin{figure}[h]
\centering
\includegraphics[scale=0.66]{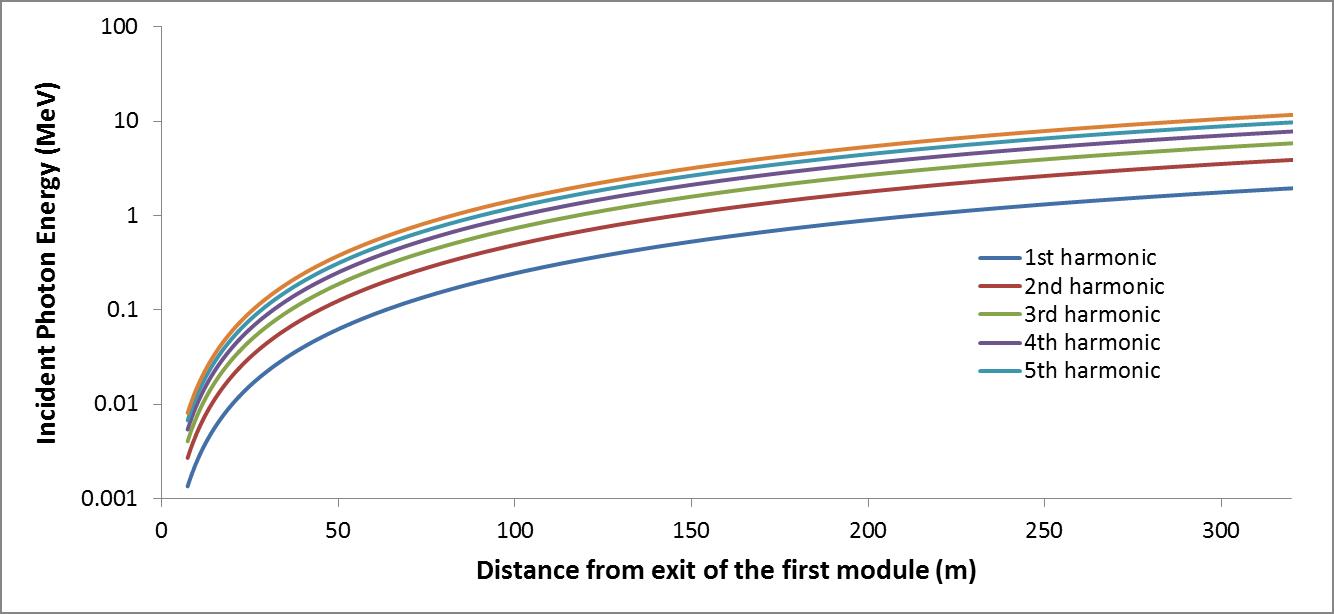}
\caption{Energy of the photons from the first module incident on the undulator walls.}
\label{fig:graph2}
\end{figure}

It is clear that the first harmonic deposits photons with lowest energy but the highest number of photons. Now we are interested in the energy deposition on the undulator wall. 

The first undulator module causes the worst scenario of power deposition on the undulator wall. The power deposition from the first module is calculated analytically. But with the analytical approach it is almost impossible  to take into account uncertainties of the undulator spectrum due to errors in period an K value.  Thus, also a simulation with the Helical Undulator Synchrotron Radiation (HUSR) code is performed. HUSR is a C++ code which was developed by David Newton on Cockcroft Institute \cite {newton2010rapid,newton2010modeling}. Here, both approaches consider an ideal undulator to verify the agreement of the results. In later studies the HUSR code will be used including realistic undulator errors.

Due to SR the electron beam loses 3 GeV along the helical undulator; thus a 128 GeV electron beam is chosen to be tracked through the undulator. The ILC undulator parameters used in this study are given in table 1. The positions of the undulator modules are taken from the ILC TDR \cite{adolphsen2013international}.

Figure 3 shows the power deposition on the wall from photons generated in the first module. Comparing the results of the analytical and simulation (HUSR) approach one finds a good agreement for distances above 30m from the exit of the first undulator module. For distances below 25m from the exit of a module, HUSR was found too time consuming to calculate the photon spectrum.

Figure 3 shows the energy deposition from the first module along the undulator – determined using the analytical approach and HUSR, and it shows the total energy deposition taking into account the SR from all 132 undulator modules. The latter was calculated analytically.

 The red line represents the total power deposition in the wall considering all 132 undulator modules.

\begin{figure}[h]
\centering
\includegraphics[scale=0.62]{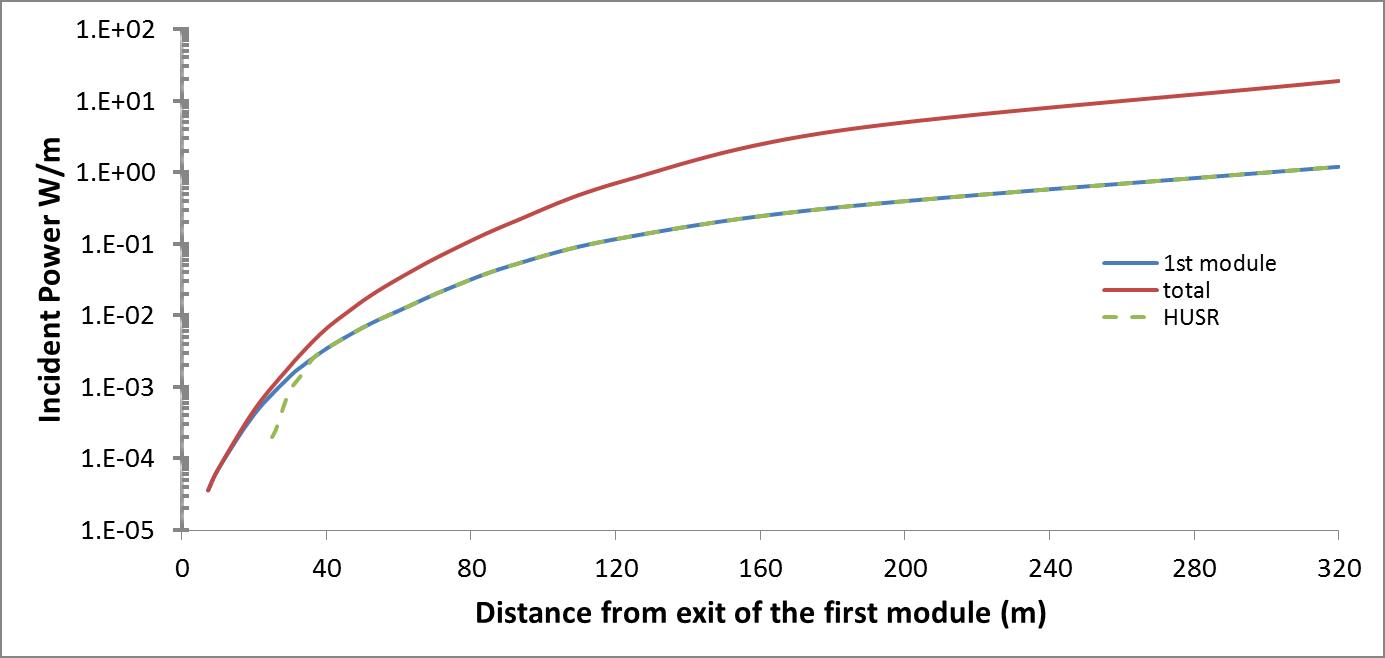}
\caption{Energy deposition in the undulator walls. The blue line represents the deposition from photons generated in the first module calculated using Equation (1) and HUSR (green dashed line); the red line shows the total energy deposition taking into account all modules.}
\label{fig:graph3}
\end{figure}

\section{Photon Masks}

Masks will be installed to protect the undulator walls from synchrotron radiation and to keep the power deposition in the wall below 1 W/m \cite {scott2008investigation}.

The peak power deposited per meter from the first module is $\approx$ 1.10 W and from all modules $\approx$ 18.5 W. This power must be absorbed by placing photon masks along the undulator length. The photon masks have a smaller aperture than the undulator aperture. In the ILC TDR, photon masks with 4.4 mm aperture have been chosen. These masks are placed behind the quadrupoles as shown in figure 4. In total, 23 photon masks are placed along the undulator; the distance between two masks is 14.538m.

\begin{figure}[h]
\centering
\includegraphics[scale=1]{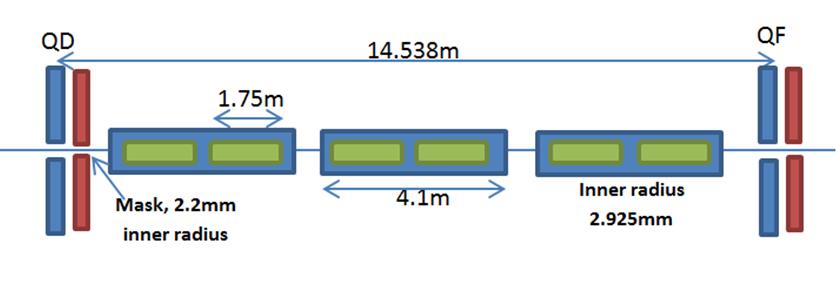}
\caption{Schematic layout of a half cell of the undulator line.}
\label{fig:graph4}
\end{figure}

If we assume that masks are ideal and absorb all power that hit them, figure 5 and figure 6 show the load on the walls at the end of the undulator, i.e. between masks 22 and 23.  Figure 5 presents the average energy and figure 6 presents the total number photons incident on the walls as function of the distance between masks 22 and 23. Figure 7 shows the total incident power on the undulator wall between these two masks.

\begin{figure}[h]
\centering
\includegraphics[scale=0.8]{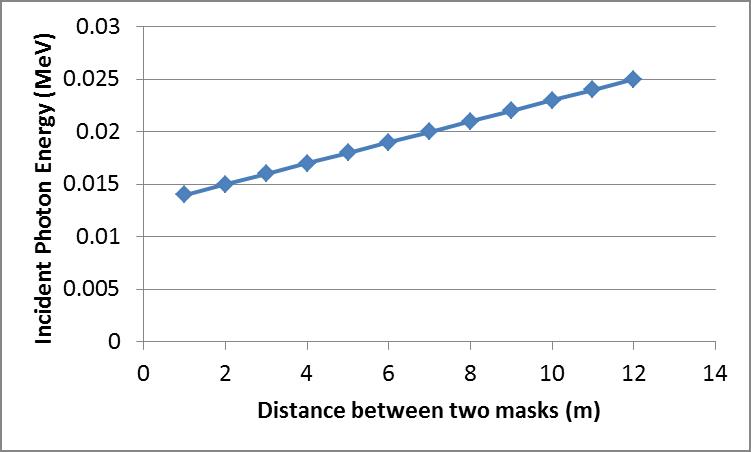}
\caption{Average photon energy incident between masks 22 and 23.}
\label{fig:graph5}
\end{figure}

The effect of adding masks on the incident power along 23 photon masks is shown in figure 8. The green points represents the location of the photon masks. Using photon masks will reduce the peak incident power to $\approx$ 0.022 W/m.

\begin{figure}[h]
\centering
\includegraphics[scale=0.8]{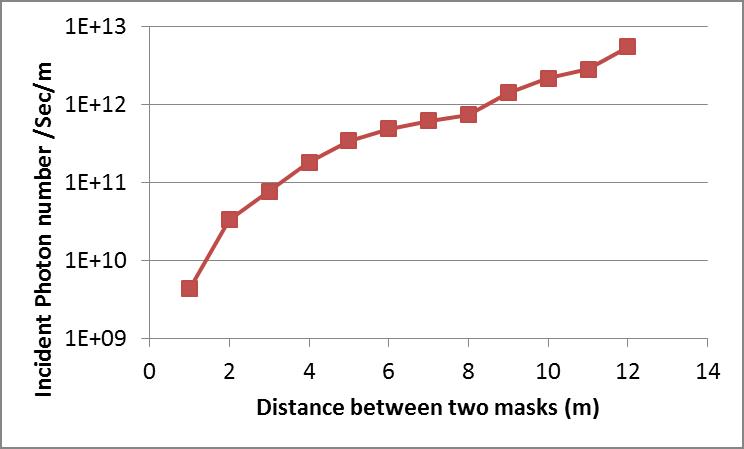}
\caption{Number of photons incident between masks 22 and 23.}
\label{fig:graph6}
\end{figure}

\begin{figure}[h]
\centering
\includegraphics[scale=0.6]{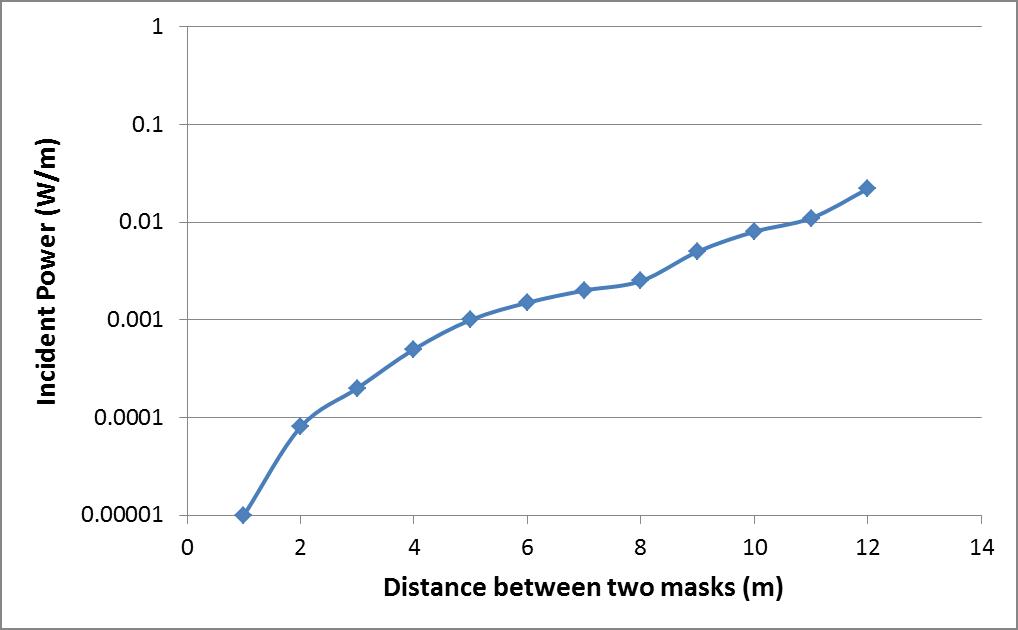}
\caption{Power deposited between masks 22 and 23.}
\label{fig:graph7}
\end{figure}

\begin{figure}[h]
\centering
\includegraphics[scale=0.6]{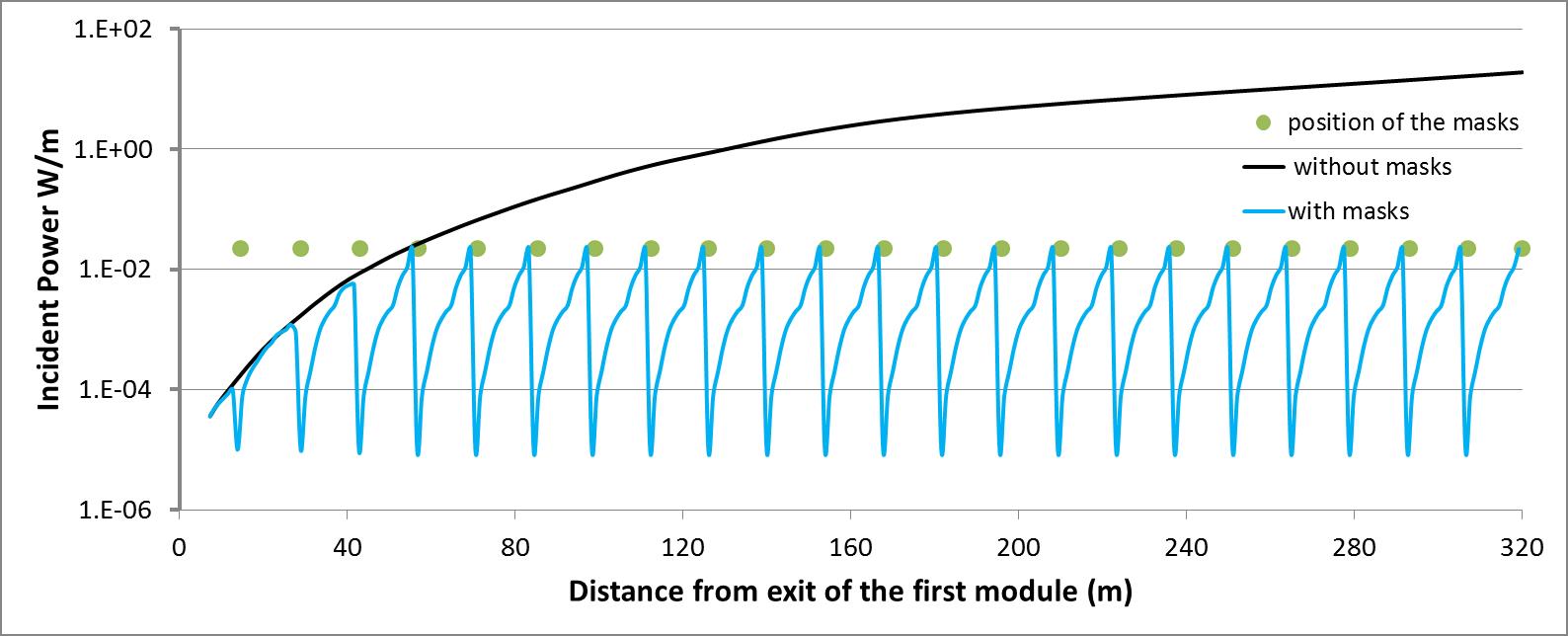}
\caption{Power incident on the undulator walls from all 132 undulator module, with masks (blue) and without masks (black). The green points represent the position of the photon masks.}
\label{fig:graph8}
\end{figure}

\section{Photon Power incident on the masks}

As shown in figure 8, photon masks with aperture of 4.4mm need to be placed along the undulator to keep the power deposition in walls below the acceptable limit. To design photon masks the photon energy distribution at the mask must be known.

Figure 9 shows the maximum and average energy of photons incident on the masks. The maximum energy of photons which hit the first mask is in the KeV range while at the last mask the maximum photon energy is in the multi-MeV range.

\begin{figure}[h]
\centering
\includegraphics[scale=0.75]{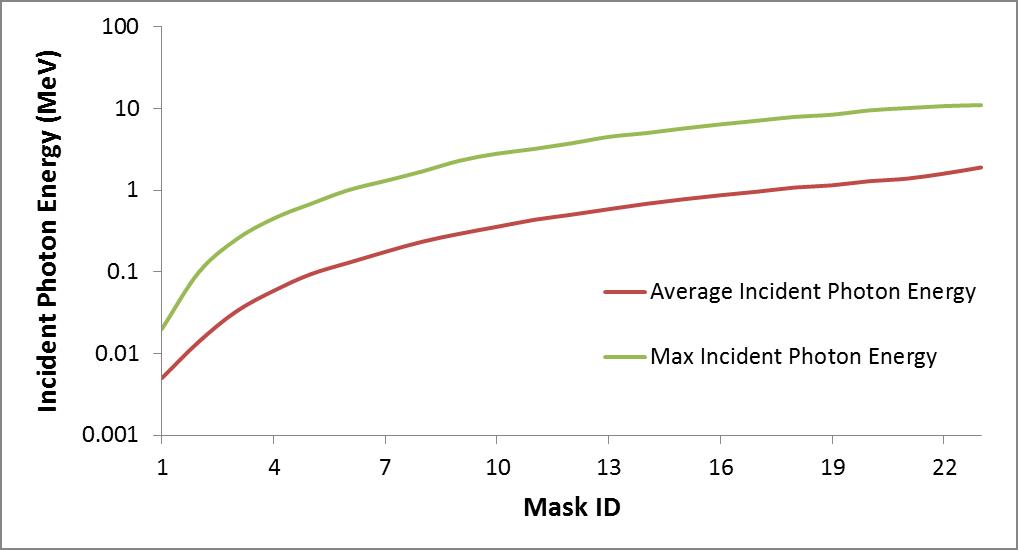}
\caption{Maximum and average energy of photons incident on the masks.}
\label{fig:graph9}
\end{figure}

At the last photon mask the peak and average energy of the photons is $\approx$ 11 MeV and $\approx$ 1.93 MeV, respectively. Figure 10 shows the energy spectrum of photons incident on the last photon mask. Although the number of photons with energies above 4 MeV drastically decreases, simulations with masks have to confirm that the masks can absorb the photons and prevent that secondaries pollute the vacuum.

\begin{figure}[h]
\centering
\includegraphics[scale=0.55]{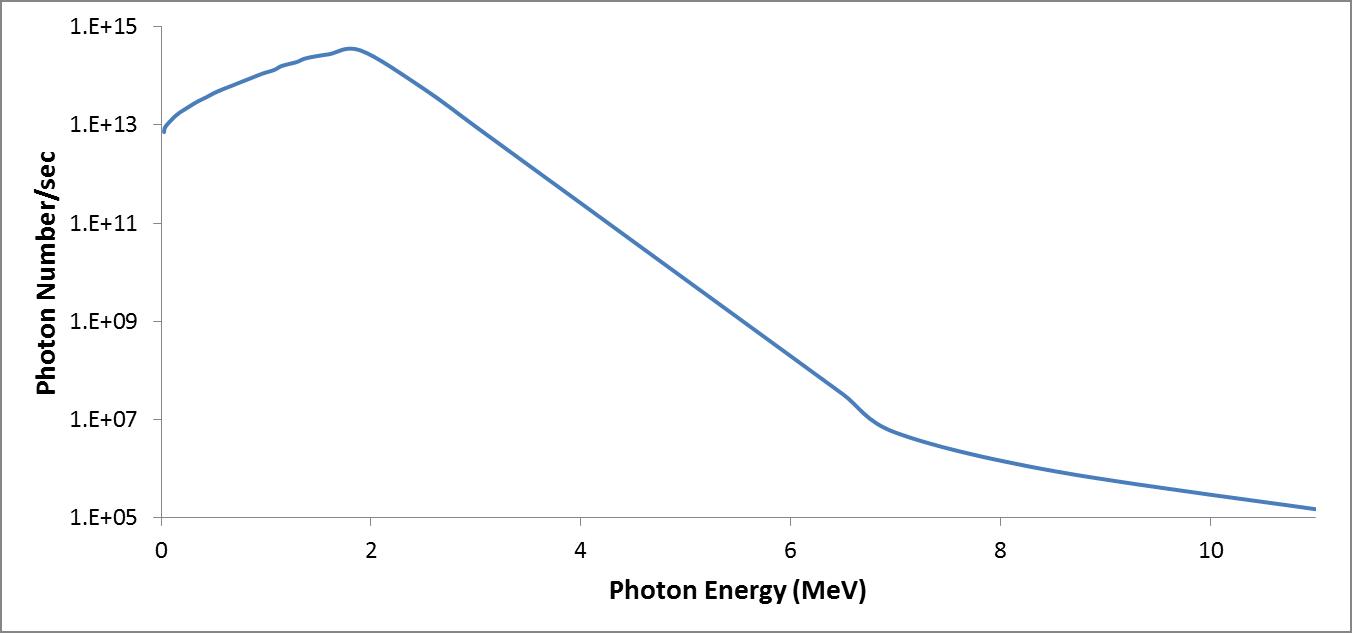}
\caption{Energy spectrum of photons incident on the last photon mask.}
\label{fig:graph10}
\end{figure}

The power incident on the last mask is shown in figure 11. As mentioned earlier, the undulator aperture is 5.85 mm and that of the masks is 4.4mm. The plot in figure 11 shows the distribution of the incident between the radii 2.2 mm and 2.925 mm. The peak deposited power is found for r = 2.2 mm, and the deposited power dramatically decreases when the radius increases to 2.925 mm, as expected. The total incident power at the last photon mask is  $\approx$ 335 W.

\begin{figure}[h]
\centering
\includegraphics[scale=0.7]{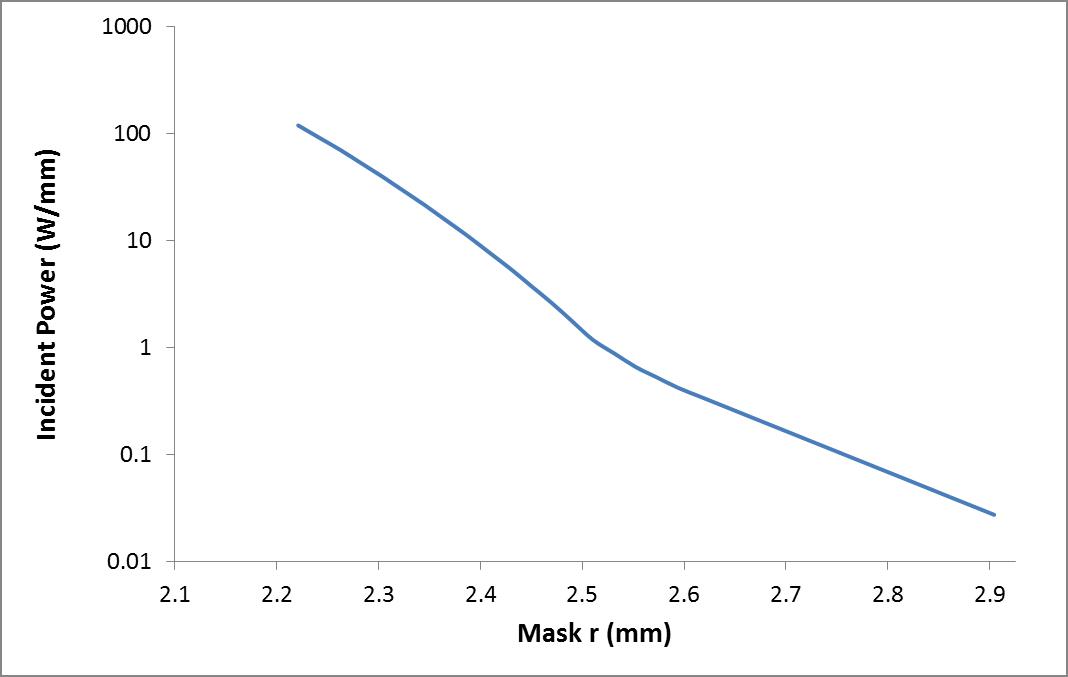}
\caption{Radial distribution of the power incident on the last mask.}
\label{fig:graph11}
\end{figure}

The energy distributions of photons on the undulator walls and photons masks are very important to design the photon mask in terms of the material and dimensions of the mask. It is clear that photon masks, in particular, which are placed at the second half of the undulator must be carefully designed to absorb all photons.

\section{Conclusion}

The power deposited in the undulator wall due to the synchrotron radiation has been studied. It is found that the peak power deposited from the first module is $\approx$ 1.1 W/m and the peak power deposited from the entire 132 modules is $\approx$ 18.5 W/m. 

Photon masks with aperture of 4.4 mm need to be inserted along the length of the undulator line to keep the power deposition in the undulator walls below the acceptable limit of 1 W/m. By adding the photon masks the peak power deposition in the undulator wall is reduced to $\approx$ 0.022 W/m.

Since we will not be able to build an ideal helical undulator the non-ideal photon distribution should be studied. This can be done by using HUSR code. The design of the 23 masks will depend on the non-ideal photon energy distribution at each mask. Therefore the non-ideal photon energy distribution at each mask will be studied.   

In this study it is assumed that all the incident power is perfectly absorbed by the photon masks. Future studies will also include the  possibility that the photon power is not completely absorbed.

\bibliographystyle{vancouver}
\bibliography{References}

\begin{thebibliography}{1}

\bibitem{adolphsen2013international}
Adolphsen C.
\newblock The International Linear Collider Technical Design Report-Volume 3.
  II: Accelerator Baseline Design.
\newblock Argonne National Lab.(ANL), Argonne, IL (United States); Thomas
  Jefferson~…; 2013.

\bibitem{flottmann1993investigations}
Flottmann K.
\newblock Investigations toward the development of polarized and unpolarized
  high intensity positron sources for linear colliders; 1993.

\bibitem{malyshev2007vacuum}
Malyshev O, Scott D, Bailey I, Barber D, Baynham E, Bradshaw T, et~al.
\newblock Vacuum Systems for the ILC helical undulator.
\newblock Journal of Vacuum Science \& Technology A: Vacuum, Surfaces, and
  Films. 2007;25(4):791--801.

\bibitem{scott2008investigation}
Scott DJ.
\newblock An Investigation into the Design of the Helical Undulator for the
  International Linear Collider Positron Source.
\newblock University of Liverpool; 2008.

\bibitem{bungau2008design}
Bungau A, Bailey I, Baynham E, Bradshow T, Brummitt A, Carr F, et~al.
\newblock Design of the Photon Collimators for the ILC Positron Helical
  Undulator.
\newblock EUROTeV-Report-2008-047. 2008;.

\bibitem{kincaid1977short}
Kincaid BM.
\newblock A short-period helical wiggler as an improved source of synchrotron
  radiation.
\newblock Journal of Applied Physics. 1977;48(7):2684--2691.

\bibitem{newton2010rapid}
Newton D, et~al.
\newblock The rapid calculation of synchrotron radiation output from long
  undulator systems.
\newblock Proceedings of IPAC2010, Kyoto, Japan. 2010;.

\bibitem{newton2010modeling}
Newton D, et~al.
\newblock Modeling synchrotron radiation from realistic and ideal long
  undulator systems.
\newblock Proceedings of IPAC2010, Kyoto, Japan. 2010;.

\end{thebibliography}

\end{document}